\documentclass[twocolumn,twoside]{ncnsd}
  \usepackage{graphicx,psfrag}
  \def\BibTeX{{\rm B\kern-.05em{\sc i\kern-.025em b}\kern-.08em
      T\kern-.1667em\lower.7ex\hbox{E}\kern-.125emX}}

\begin{document}
  \title{Clustering of advected passive sliders on a fluctuating 
  surface} 
  \author{Apoorva Nagar, Mustansir Barma and Satya N. Majumdar}
  \thanks{A. Nagar (corresponding author)  and M. Barma are with the Department of
  Theoretical Physics, Tata Institute of Fundamental Research, Homi Bhabha road,
  Mumbai - 400005, India. Satya N. Majumdar is at Laboratoire de Physique
  Quantique, Universit'e Paul Sabatier, 31062 Toulouse Cedex, France. E-mail:
  apoorva@theory.tifr.res.in}
  \markboth{NATIONAL CONFERENCE ON NONLINEAR SYSTEMS \& DYNAMICS} 
  {INDIAN INSTITUTE OF TECHNOLOGY, KHARAGPUR 721302, DECEMBER 28-30,
  2003}
  \maketitle

  \begin{abstract}
  We study the clustering properties of advected, non-interacting,
   passive scalar particles in a Burgers fluid with noise, a problem which maps to
  that of passive sliding particles moving under gravity on a surface
  evolving through the Kardar-Parisi-Zhang equation.  Numerical
  simulations show that both the density-density correlation function
  and the single-site mass distribution scale with system size.  The
  scaling functions diverge at small argument, indicating strong
  clustering of particles. We analytically evaluate the  
  scaling functions for the two-point correlation and mass distribution of 
  noninteracting particles in thermal
  equilibrium in a random landscape, and find that the results 
  are remarkably similar to those for nonequilibrium advection.
   
  \end{abstract}

  \section{Introduction}
  The advection of a passive scalar in a turbulent fluid is an
  interesting problem which has, in the recent past, attracted
  considerable attention~\cite{kraichnan}~\cite{shraiman}~\cite{falkovich}.  By a
  passive scalar we mean a scalar field (e.g. temperature or density
  of dye) that is advected by the fluid flow but which has no back
  effect on the flow.  In this paper we will consider the problem of a
  passive particle density in a compressible fluid described by the
  Burgers equation in the presence of white noise, studied recently in~\cite{drossel}.  While an 
  incompressible density
  field would tend to disperse and ramify when carried by an
  incompressible flow, the reverse can happen if the fluid is
  compressible, and particle clustering (rather than dispersal) can
  result.  The characterization of this clustering in the
  one-dimensional case is the principal objective of this paper.

  The restriction to flows described by the Burgers
  equation~\cite{burgers}~\cite{jayaprakash} allows the problem to be mapped to another
  interesting problem, namely that of passive, sliding particles (called
  sliders) moving under gravity along a growing surface which is itself
  fluctuating in time.  Previous studies of the related problem of
  hard-core particles sliding on a fluctuating surface have shown the
  occurrence of a new and interesting state, with large-scale clustering
  of particles~\cite{dasbarma}~\cite{dasbarmasatya}. Now
  in the problem under consideration here, there is no hard-core
  interaction between sliders, so we may expect even stronger clustering
  properties.  As discussed below, this expectation is borne out in our numerical simulations of the system. 
  There is a clear signature of clustering in the behaviour of the
  density-density correlation function $C(r)$, which has a scaling form
  with argument $r/L$ where $L$ is the system size.  While in the 
  fluctuation-dominated phase ordering (FDPO)
  state discussed in ~\cite{dasbarma} and~\cite{dasbarmasatya}, the
  scaling function has a {\it cusp} singularity at small argument, we find
  that it has a {\it divergence} in the problem under study here.
  Interestingly, we find that this strongly nonequilibrium system has
  the same scaling properties as a system of noninteracting particles in
  thermal equilibrium in a random, static landscape. The latter problem is
  solved analytically, and the results are found to be similar to the numerical results for 
  nonequilibrium    advection.
  \bigskip

  \section{Model}

  We first discuss the continuum equations describing the passive scalar
  problem with Burgers flow. We then describe a one dimensional                  lattice model which is
  expected to have the same scaling properties.

  The velocity field $\vec v$ of a randomly stirred Burgers fluid is
  described by the equation 
  \begin{equation} 
  {\partial \vec v \over \partial t} + \lambda(\vec v \cdot \nabla) 
  \vec v = \nu \nabla^2 \vec v + \nabla \zeta_h (\vec x,t) 
  \label{one} 
  \end{equation} 
  Here $\nu$ is the viscosity, while the random stirring is caused by
  Gaussian white noise $\zeta_h$ satisfying $\langle \zeta_h (\vec
  x,t) \zeta_h({\vec x}',t')\rangle = 2D_h \delta^d(\vec x - {\vec x}')
  \delta(t - t')$.

  A passive scalar particle moving with the flow follows the equation
  \begin{equation} 
  {d\vec x \over dt} = a\vec v + \zeta_x (t)
  \label{two} 
  \end{equation} 
  where the white noise $\zeta_x (t)$ represents the randomising effect
  of temperature, and satisfies $\langle \zeta_x (t) \zeta_x(t')\rangle
  = 2\kappa \delta(t - t')$.  The parameter $a$ governs the coupling of
  the particle to the flow.  If the flow is vortex-free, we may write
  $\vec v = -\nabla h$, where the field $h$ satisfies the equation
  \begin{equation} 
  {\partial h \over \partial t} = \nu \nabla^2 h + {\lambda \over 2} 
  (\nabla h)^2 + \zeta_h(\vec x,t) 
  \label{three} 
  \end{equation} 
  This is the Kardar-Parisi-Zhang (KPZ)~\cite{kpz} equation for surface
  evolution with $h(\vec x,t)$ describing the height of a growing
  surface at position $\vec x$ at time $t$.

  Rather than analysing the coupled Eqs. (2) and (3) directly, we will study a lattice model  which is expected to have similar behaviour at large length and time scales. 
  The model consists of a flexible one-dimensional lattice in
  which particles reside on sites, while the links or bonds between
  successive lattice sites are also dynamical variables denoting local
  slopes.  Each link takes values $+1$ (upward slope $\rightarrow /$)
  and $-1$ (downward slope $\rightarrow \backslash$).  We evolve the
  surface by choosing a site at random, and if it is on a local
  hill$(\rightarrow /\backslash)$, the local hill is changed to a local valley
  $(\rightarrow \backslash /)$.  These moves of the links define the well
  known single-step model, whose large distance, long time properties
  are known to be identical to those of the KPZ
  equation.  The particles (any number of which can be present at a
  given site) are moved one at a time.  We choose a particle at random
  and move it one step downward with probability $(1+K)/2$ or upward
  with probability $(1-K)/2$.  The parameter $K$, which ranges from 1
  (particles totally following the surface slope) to 0 (particles moving
  independently of the surface) mimic the ratio $a/\lambda$ in Eqs. (2)
  and (3).  In our simulations, we update the surface and particles at
  independent sites, reflecting the independence of the noises $\zeta_h
  (\vec x,t)$ and $\zeta_x (t)$.  This is in contrast to Kardar and
  Drossel's update rule~\cite{drossel} where only particles residing at a
  site affected by the surface evolution are moved.

  \begin{figure}
  \centering
  \includegraphics[width=0.7\columnwidth,angle=-90]{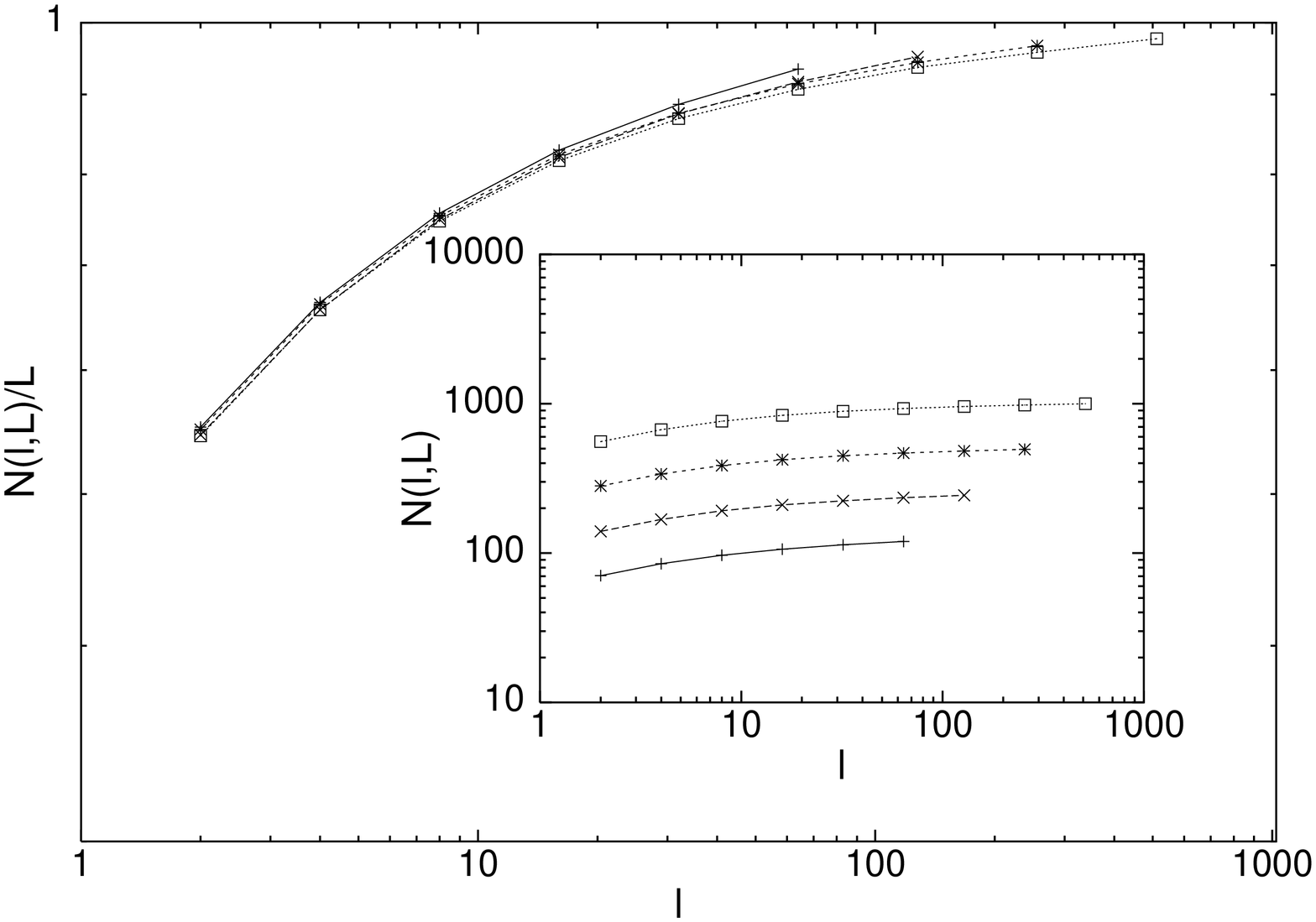}
  \caption{The number N(l,L) of particles which lie in the same
  bin of size l as a randomly chosen particle is shown in the inset for various system sizes L. The various sizes are 128, 256,512, 1024 as we move up. The main figure shows the scaling of N(l,L) with L}
  \label{fig1label}
  \end{figure}

  \section{Numerical Results}
   
  As particles move preferentially downwards, it is evident that they 
  tend to move towards valleys, and this valley-seeking tendency 
  promotes clustering. The question is how one can quantify its nature
  and extent. The numerical study of the two-point 
  correlation function and single-site particle distribution described in this section show that a scaling description affords a simple and compact description of clustering properties. 

  We consider the steady state of a system of $M$ particles in a system with $L$ sites.  In
  our numerical study, we used $L$ in the range 128 to 1024, and took $M
  = L$. The successive update times of the particle and the surface
  moves were taken to be equal to each other.
  To start with, we monitored the RMS displacement $x_{RMS}$ of a given
  tagged particle, and verified that $x_{RMS} \sim t^{1/z}$ where $z =
  3/2$ is the KPZ value of the dynamic exponent.  This is expected on
  the basis that the particle is likely to be in the largest fresh 
  valley that has formed within time $t$ in the
  vicinity of the particle, as such valleys have a spatial size of 
  order $t^{1/z}$.  Quite
  direct evidence of particle clustering is obtained by choosing a
  particle at random and monitoring the mean number of particles within
  a distance $\ell$ of it.  The results are shown in Fig. 1.  Saturation
  of the curves as $\ell$ increases is a clear indication of particle
  clustering.  These results are essentially identical to those of
  Drossel and Kardar~\cite{drossel} for their slightly different model.

  Our results for the two-point (unconnected) density-density
  correlation function $G(r,L) = \langle m_i m_{i+r}\rangle$ for
  different $L$ are shown in Fig. 2, where $m_i$ is the number of particles at   site $i$. There is strong evidence that the
  scaling form 
  \begin{equation} 
  G(r,L) \sim L^{-\theta} Y(r/L) 
  \end{equation}
  is valid with $\theta = {1\over2}$, and that the scaling
  function $Y(y)$ has a power law divergence $Y \sim y^{-\phi}$ as $y
  \rightarrow 0$. The value of $\phi$ is close to 1.5, which is the value 
  obtained for particles which have reached equilibrium on a random surface, as  discussed in section IV below.

  The divergence of the scaling function at small argument is the result
  of strong clustering. It is an outcome of the unrestricted occupancy
  of a site by noninteracting particles, and quite different from the
  milder cusp singularity found in the FDPO state with hard-core
  exclusion~\cite{dasbarmasatya}. As shown in Section IV, the form of the scaling function can be found analytically for the ostensibly different problem of noninteracting
  particles in thermal equilibrium on a random landscape. As a matter of
  fact, the analytic forms describe quite well the data in Fig. 2 
  for the strongly nonequilibrium system under study.
  This is discussed further in Section V.

  Another interesting quantity, which however does not carry any spatial
  information, is the probability $P (m,L)$ that any site has occupancy
  $m$. As is evident from Fig. 3, $P(m,L)$ follows the scaling form
  \begin{equation}
  P(m,L) \sim {1\over L^2} f \left({m\over L}\right). 
  \end{equation}
  At first sight, the scaling function seems to follow a power law $f(y) \sim y^{-\tau}$ with $\tau$ close to 1, but guided by the analytic theory to be discussed in the next section, we find that the $y \rightarrow 0$ behaviour is well described by the form $f(y) \sim y^{-1} \ln (b/y)$ with $b \simeq 5$. On
  integrating over $m$, we find $1-P(0,L) \sim \frac{1}{L}(\ln L)^2$,
  implying that the number of occupied sites increases with size as
  $(\ln L)^2$. This indicates that the particles are distributed 
  within a relatively
  compact region in space. Finally, we note that Eq.(5) leads to 
  $\langle m^2 \rangle \equiv G(0,L)\sim L$. Direct simulation results for 
  $G(0,L)$ verify the linear dependence on $L$, lending further support to the 
  scaling form (Eq.(5))

  \section{Analytical Results (Equilibrium)}
Since the particles are noninteracting in our model, the number density
$\rho(r,t)=m(r,t)/L$ of particles is identical to the probability that
a single particle, moving in the fluctuating landscape, will be at a position 
$0\leq r \leq L$ at time $t$. Thus it suffices to consider one single particle
moving in the fluctuating landscape. We consider the limiting 
case when changes in
the surface occur over a time scale much larger than the time scale of 
fluctuations of the particle. In this limit, for any given surface configuration $h(r)$,
the Langevin equation (Eq.(2)) describes the evolution of a particle in a potential $h(r)$.
Before the surface configuration $h(r)$ can change, the particle will relax to its
its equilibrium Gibbs state, where the probability of finding the particle at position $r$
is given by $\rho(r)= \exp[-\beta h(r)]/Z$ with the partition function $Z=\int_0^L \exp[-\beta 
h(r')]dr'$, where $\beta=1/{k_B T}$. Clearly $\rho(r)$ is a random variable which changes from one 
configuration
of $h(r)$ to another. We assume that the surface itself has reached the stationary state of Eq. (3).
It is well known that for the $1$-d KPZ equation, the stationary state is described by the following 
measure, ${\rm Prob} [\{h(r\}]\propto \exp\left[-{1\over {2}}\int h^2(r')dr'\right]$. Thus, any 
stationary 
configuration can be thought of as the trace of a random walker in space evolving via
the equation, $dh(r)/dr = \xi(r)$ where the white noise $\xi(r)$ has zero mean and is
delta correlated, $\langle \xi(r)\xi(r')\rangle = \delta (r-r')$. 
Thus our problem then reduces precisely to the celebrated Sinai model
~\cite{sinai} where a particle
moves in a random potential which itself is a random walk in space. There is a slight
difference however. The periodic boundary condition on the surface, $h(r)=h(r+L)$, indicates
that we are considering a random potential which is pinned at its two ends, i.e. a
Brownian bridge, rather than a free walk.

Given this stationary measure of the 
surface configurations, we then want to evaluate the correlation function $\langle 
\rho(r_0)\rho(r+r_0)\rangle= G(r,L)/L^2$ (evidently independent of $r_0$ due to translational 
invariance) where the angular brackets denote the average over the surface configurations 
sampled from the stationary measure mentioned above. Thus we have
\begin{equation}
L^{-2} G(r,L)= \langle \left[{ {e^{-\beta [h(r_0)+h(r_0+r)]}}\over {Z^2}}\right] \rangle.
\label{eq1}
\end{equation}
Fortunately, the object on the right hand side of Eq.(\ref{eq1}) was evaluated exactly
by Comtet and Texier~\cite{comtet} in the completely different context of one dimensional
disordered supersymmetric quantum mechanics, where the right hand side of Eq. (\ref{eq1})
is simply the correlation function in the ground state wave function. They actualy evaluated 
the $n$-point correlation function. Adapting their results to our case with $n=2$, we find the
following expression
\begin{eqnarray}
G(r,L)~= ~\sqrt{2\pi} { {(\beta^2 L)^{5/2}}\over {256}}
\int_0^{\infty}\int_0^{\infty} dk_1dk_2 k_1k_2 
(k_1^2-k_2^2)^2 \\ 
\times \nonumber{ {\sinh(\pi k_1)\sinh(\pi k_2)}\over {[\cosh(\pi k_1)-\cosh(\pi k_2)]^2}} \exp\left[-{{\beta^2}\over {8}} \left( k_1^2(L-r)+k_2^2 r \right) \right].
\label{eq2}
\end{eqnarray}
Note that the expression for $G(r,L)$ has the expected symmetry, $G(r,L)=G(L-r,L)$. It is easy to
evaluate $G(0,L)$ from Eq. (\ref{eq2}) for large $L$. We find, $G(0,L)\approx \beta^2 L/{12}$ for
large $L$. On the other hand, in the scaling limit, $r\to \infty$, $L\to \infty$ but keeping the ratio 
$y=r/L$ fixed, one finds from Eq. (7) that for $0<x<1$,
$G(r,L)\sim L^{-1/2} Y(r/L)$ where the scaling function is given by
\begin{equation}
Y(y) = {1\over {\beta \sqrt{2\pi}}} [y(1-y)]^{-3/2}.
\label{scaling}
\end{equation}
Note that the point $r=0$ is not part of the scaling function. In fact, the power law 
divergence of the scaling function as $y\to 0$, $Y(y)\sim 
y^{-\phi}$ with $\phi=3/2$, is necessary in order that $G(r\to 0, L)\sim L$ for large $L$.\\
 
This formalism can also be used to calculate the equilibrium probability density $P(\rho,L)$. Details of the calculation will be given elsewhere. Our results indicate that $P(\rho,L)$ which can be written as the sum of two parts :

\begin{equation}
P(\rho,L) \approx \left[1- {{\ln^2 (L)}\over {\beta^2 L}}\right]\delta(\rho) + 
{4\over {\beta^4 L}} G\left[ {{2\rho}\over {\beta^2}}\right]\theta\left(\rho-{c\over {L}}\right),
\label{scaled2}
\end{equation}

The first part refers to vacant stretches, and to the fact that the number of occupied sites occupies a vanishing fraction $\sim (\ln L)^2/L$ of the system. The scaling function $G(y)$ in the second part is given by

\begin{equation}
G(y) = { {e^{-y}}\over {y} } K_0(y).
\label{gy1}
\end{equation}
where $K_0(y)$ is the modified Bessel function which has the asymptotic behaviour $[-\ln (y/2)-0.5772...]$ as $y \rightarrow 0$. The theta function incorporates a lower cutoff on the validity of the scaling form.\\ 
                                                               
The equilibrium-based results obtained in this section appear to describe rather well the behaviour of the nonequilibrium system studied numerically in section III. For instance Eq.(8) describes the behaviour (Eq.(4)) of the two-point correlation function (Fig. 2), with $\beta \simeq 4$ giving good agreement. The data for the mass distribution (Fig. 3) is quite well described by Eq.(10), but with a different value of $\beta$ ($\simeq 2.3$). An intermediate value of $\beta (\simeq 2.7)$ gives tolerable, but not very satisfactory, agreement with both sets of data (Figs. 2 and 3).

 \begin{figure}
  \centering   
  \includegraphics[width=0.7\columnwidth,angle=-90]{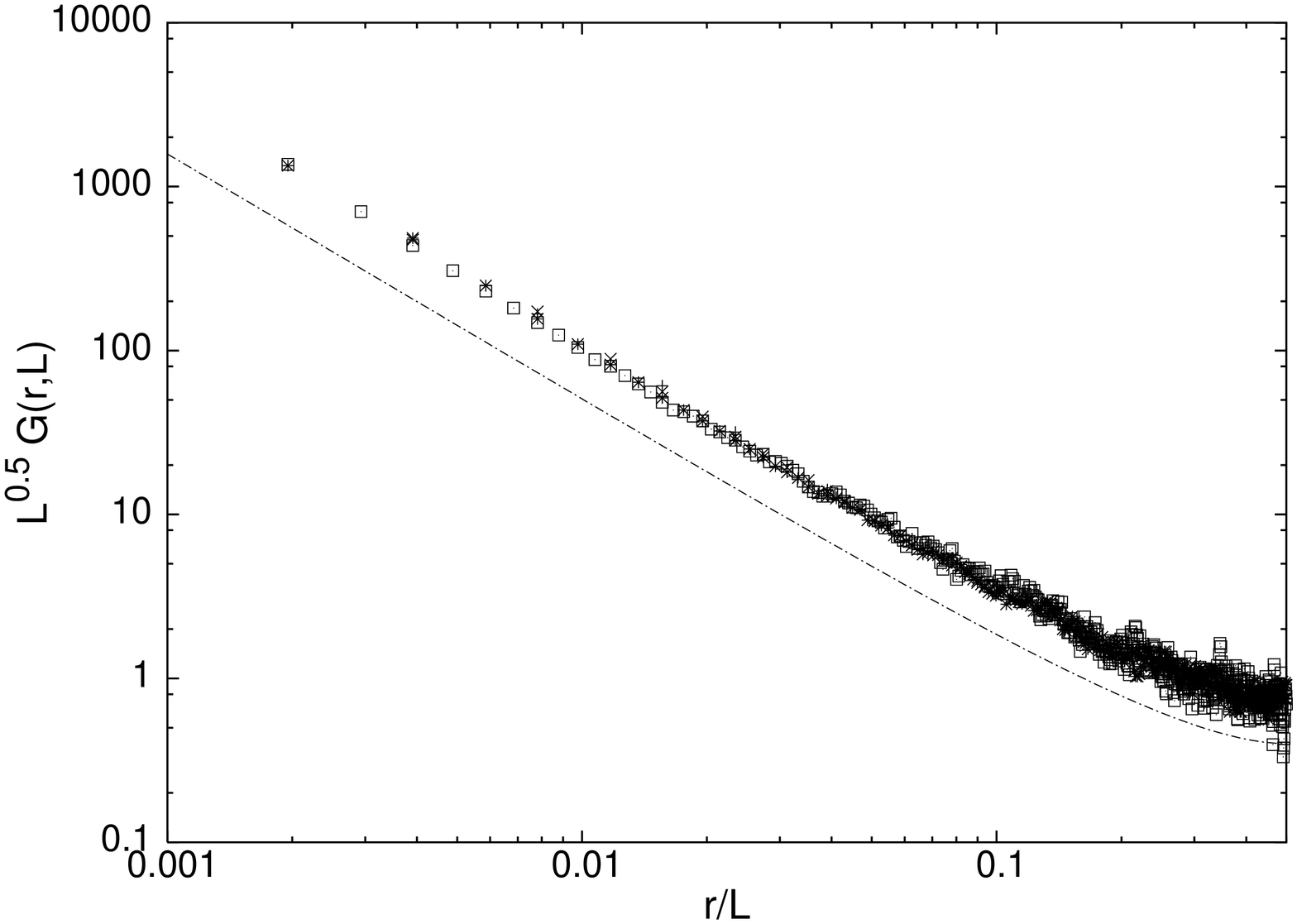}
  \caption{Scaling of the density-density correlation function G(r,L) with 
   system size L. The dashed curve is a plot of the scaling function Y (Eq.(8))    shifted downwards for clarity.}
  \label{fig1label}
  \end{figure}

  \begin{figure}
  \centering
  \includegraphics[width=0.7\columnwidth,angle=-90]{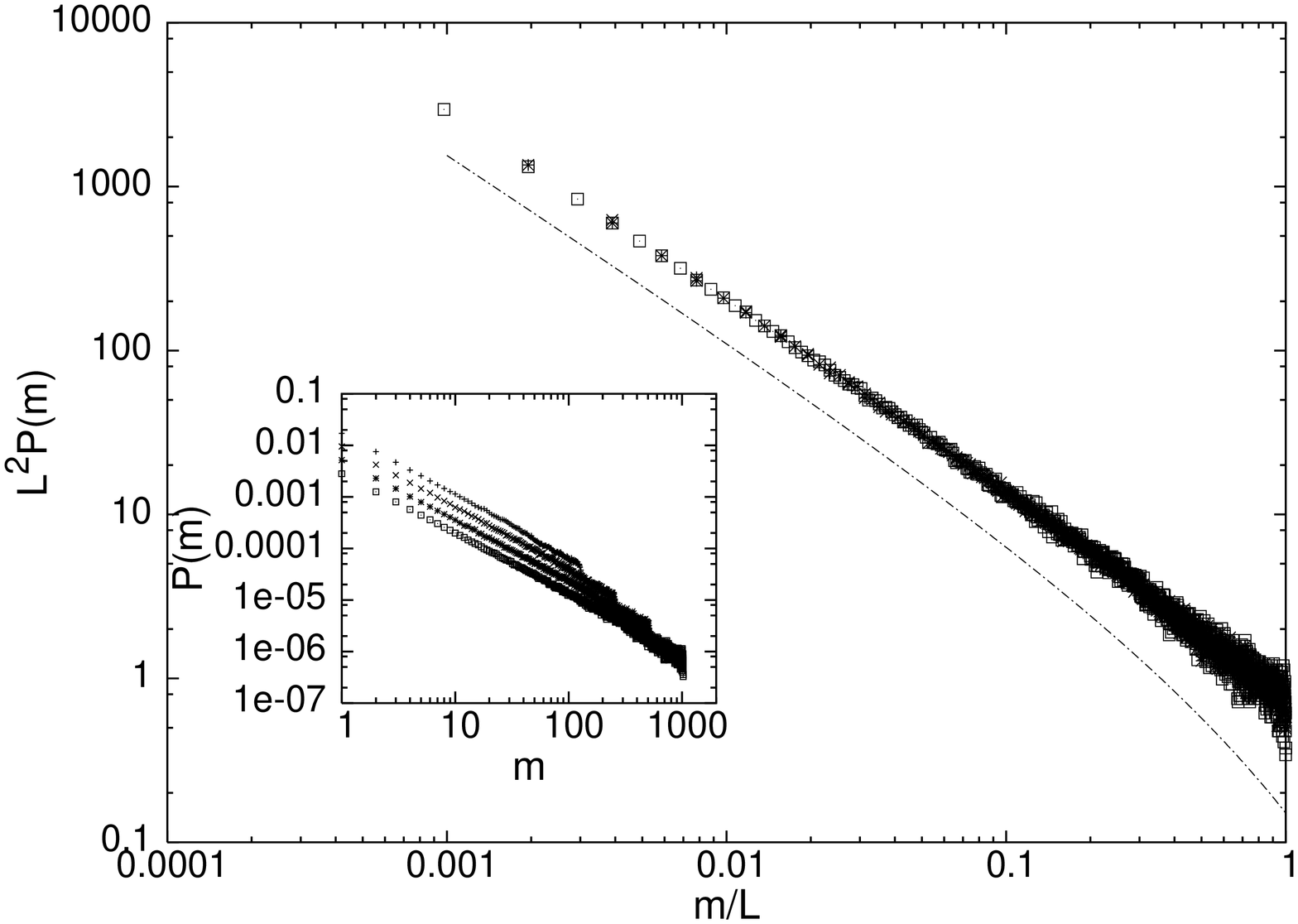}
  \caption{The probability  P(m,L) of finding a site containing m particles is shown in the inset for various system sizes L. The main figure shows the scaling of P(m,L) with L; the dashed curve shows the scaling function G  (Eq.(10)) shifted downwards for clarity. }
  \label{fig1label}
  \end{figure}

  \section{Discussion}

        As we have seen in the previous two sections, a number of the
properties of the nonequilibrium system of particles sliding in a
fluctuating landscape are strikingly similar to those of an equilibrium
system of particles settling in a quenched random landscape. This raises
two questions: First, what is the source of this similarity in
behaviour? And second, how robust is it? We address these
below.\\

        In thermal equilibrium, in a given configuration of the landscape,
the Boltzmann factor guarantees that most particles are to be found
within a height $T$ of the global minimum, where $T$ is the temperature. In the nonequilibrium situation,
by contrast, examination of several typical configurations shows that
particles cluster near valley bottoms, but these valleys are often not the
globally
lowest one. Why then are properties similar when averaged over
configurations? A partial  answer lies in the fact that terrain around {\it
any}
reasonably deep valley is statistically similar. This feature of the
`random walk' landscape under consideration can plausibly lead to particles
being distributed in similar fashions in the two cases. \\

Evidently, this happens despite the fact that the character of the noise which
leads to particle re-settling is quite different in the two cases: in the
equilibrium problem, particles are acted upon by thermal noise, while in the
nonequilibrium
case, it is surface fluctuations which drive re-settling.
Thus, despite the similarities in the landscape, it is not clear to what
extent one should expect an equivalence of results, and indeed, as the
comparison in Section IV shows, the equivalence for different properties
involves equilibrium systems at different temperatures.  It would also be interesting to see how robust these results are with respect to variations of the parameter $\omega$, which is the ratio of the times $\tau_s$ between successive
updates of the surface, and $\tau_p$ between successive updates of the
particles. In the studies reported in this paper, $\omega$ was held
constant at 1. Preliminary investigations suggest that variations of
$\omega$ can induce interesting features specific to the nonequilibrium
problem.\\

\begin{center}

ACKNOWLEDGEMENT

\end{center}
S.N.M. acknowledges useful discussions with A. Comtet.




  \bibliographystyle{ncnsd}

\begin{thebibliography}{1}
  \bibitem{kraichnan}
  R.H. Kraichnan, ``Anomalous scaling of a randomly advected passive scalar'',\newblock{\em Phys. Rev. Lett.}\newblock vol. 85, pp. 1016-1019 (1994)
  \bibitem{shraiman}
  B. I. Shraiman and E. D. Siggia, ``Scalar turbulence'',\newblock{\em Nature} \newblock vol. 405, pp. 639-646 (2000)
  \bibitem{falkovich} G. Falkovich, K. Gawedzki, and M. Vergassola, ``Particles and fields in fluid turbulence'',\newblock{\em Rev. Mod. Phys.}\newblock vol.73, pp. 913-975 (2001)
  \bibitem{drossel}
  B. Drossel and M. Kardar, ``Passive sliders on growing surfaces and advection in Burgers flows'', \newblock{\em Phys. Rev. B} \newblock vol. 66, pp. 195414-1-4
  (2002)
  \bibitem{burgers}
  J. M. Burgers,\newblock{\em The non-linear diffusion equation: asymptotic
  solutions and statistical problems},\newblock Kluwer Academic Publishers, 2002  \bibitem{jayaprakash}
   F. Hayot and C. Jayaprakash, ``Aspects of the stochastic Burgers equation and their connection with turbulence'', \newblock{\em Int. J. Mod. Phys. B},\newblock vol. 14, pp. 1781-1800 (2002)

  \bibitem{dasbarma}
  D. Das and M. Barma, ``Particles sliding on a fluctuating surface: phase separation and power laws'', \newblock{\em Phys. Rev. Lett.} \newblock vol. 85, pp. 1602-1605 (2000)
  \bibitem{dasbarmasatya}
  D. Das, M. Barma and S. N. Majumdar, ``Fluctuation-dominated phase ordering driven by stochastically evolving surfaces: depth models and sliding particles'',\newblock{\em Phys. Rev. E }\newblock vol. 64, pp. 046126-1-16 (2001)
  \bibitem{kpz} M. Kardar, G. Parisi, Y. Zhang, ``Dynamic scaling of growing interfaces'', \newblock{\em Phys. Rev. Lett.} \newblock vol. 56, pp. 889-892 (1986)
  \bibitem{comtet} A. Comtet and C. Texier, ``One-dimensional
  disordered supersymmetric quantum mechanics: a brief survey'' in \newblock{\em Supersymmetry and
  Integrable Models}, Aratyn, Imbo, Keung, Sukhatme (eds.), Proceedings Chicago IL,
  1997, Lecture notes in Physics,\newblock Springer, (1998); cond-mat/9707313 \bibitem{sinai} Y.G. Sinai, ``The limiting behavior of a one-dimensional random walk in a random medium'',\newblock{\em Theor. Probab. Appl}. vol. 27, pp. 256-268 (1982).
  \bibitem{manoj} M. Gopalakrishnan, ``Anomalous diffusion of a passive sliding particle on a fluctuating surface'',\newblock{\em cond-mat/0308109}

  \end{thebibliography}

  \end{document}